\documentclass[12pt]{iopart}
\pdfoutput=1
\usepackage{amssymb}
\usepackage{amsbsy}
\usepackage{graphicx}
\usepackage{epstopdf}
\usepackage{color}
\usepackage[colorlinks,bookmarks=false,citecolor=blue,linkcolor=red,urlcolor=blue]{hyperref}

\def\be{\begin{equation}}
\def\ee{\end{equation}}

\begin{document}

\title[MC study of the 3D $E_g$ model]{Monte Carlo study of the critical properties of the three-dimensional $120^\circ$ model}

\author{Sandro Wenzel}
\address{Institute of Theoretical Physics, \'Ecole Polytechnique F\'ed\'erale de Lausanne (EPFL), CH-1015 Lausanne, Switzerland}
\author{Andreas M L\"auchli}
\address{Max-Planck-Institut f\"ur Physik komplexer Systeme, N\"othnitzer Str.\ 38,\\ D-01187 Dresden, Germany}
\address{Institut f\"ur Theoretische Physik, Universit\"at Innsbruck, Technikerstra\ss e 25/2,\\ A-6020 Innsbruck, Austria}

\begin{abstract}
We report on large scale finite-temperature Monte Carlo simulations of
the classical $120^\circ$ or $e_g$ orbital-only model on the simple
cubic lattice in three dimensions with a focus towards its critical
properties. This model displays a continuous phase transition to an
orbitally ordered phase.  While the correlation length exponent
$\nu\approx0.665$ is close to the 3D XY value, the exponent $\eta
\approx 0.15$ differs substantially from O(N) values. We also
introduce a discrete variant of the $e_g$ model, called $e_g$-clock
model, which is found to display the same set of exponents. Further,
an emergent U(1) symmetry is found at the critical point $T_c$, which
persists for $T<T_c$ below a crossover length scaling as $\Lambda \sim
\xi^a$, with an unusually small $a\approx1.3$.
\end{abstract}

\pacs{05.70.Fh, 64.60.-i, 75.10.Hk, 75.40.Mg}

\section{Introduction}
Orbital degrees of freedom are a key ingredient to the rich physics
observed in many transition metal compounds, where in combination with
magnetic and charge degrees of freedom complex phase diagrams are
realized \cite{TokuraScienceReview}. Several paradigm spin models have
been introduced to describe the physics originating from the
collective interplay of those orbital degrees of freedom. In their
most pure form, these models are called \emph{orbital-only} models
neglecting all but orbital degeneracy
\cite{KhomskiiJPhysA2003,vandenBrink_NJP}. Beyond their original
motivation, these models are also discussed in quite a different
context, e.g. in connection to quantum information
\cite{doucot:024505}. It is quite surprising, though, that despite
their prototype status only little is known about the
finite-temperature properties, and in particular about their critical
properties and the precise nature of orbital-ordering thermal phase
transitions in three-dimensions.

Here, we present results of a comprehensive Monte Carlo (MC)
investigation of the nature of the finite-temperature phase
transitions of the prototypical $120^\circ$ orbital-only model on the
three-dimensional (3D) cubic lattice, complementing on our recent
short account \cite{WenzelEg2011}. The $120^\circ$ model is the most
basic model describing $e_g$ orbital degeneracy of electrons in the
$d$-shell, hence the model is often also called the $e_g$ model.  We
study here the classical version because the corresponding quantum
model has a sign problem precluding Quantum Monte Carlo approaches,
and because in Ginzburg-Landau theory one typically expects quantum
and classical versions of a same model to have the same critical
properties, although exceptions are possible. For a detailed
finite-temperature treatment of the three-dimensional compass model --
the second major orbital-only model~\cite{vandenBrink_NJP} -- the
reader is referred to a forthcoming publication~\cite{Wenzel3DCM}.

\section{Definition of the $120^\circ$ model}
The $120^\circ$ or $e_g$ model (EgM) is defined by the Hamiltonian~\cite{vandenBrink_NJP}
\begin{equation}
    \label{eqn:eg_model}
    \mathcal{H}_{e_g}= -\ J\sum_{i, \alpha } {\boldsymbol{\tau}}_i^\alpha {\boldsymbol{\tau}}_{i+\mathbf{e}_\alpha}^\alpha,
  \end{equation}
  where $\boldsymbol{\tau}_i$ is an auxiliary three component vector obtained by
  an embedding of the orbital degree of freedom $\mathbf{T}_i=(T^z_i,T^x_i)=(\cos(\varphi),\sin(\varphi))\in S^1$:
  \be
  \label{eqn:transformation}
  {\boldsymbol{\tau}}_i = \pmatrix{ -1/2 & \sqrt{3}/2 \cr -1/2 &  -\sqrt{3}/2  \cr 1 & 0 } \mathbf{T}_i.
  \ee
  The $\boldsymbol{\tau}_i$ vector is therefore constrained onto a specific greater circle on
  $S^2$. The $\mathbf{e}_\alpha$ denote the positive unit vectors in the $\alpha
  \in \{x,y,z\}$ cartesian directions. Note that the coupling in
  $\boldsymbol{\tau}$-space depends on the spatial orientation of the
  bond. The coupling constant $J$ is set to one in the following,
  corresponding to ferromagnetic interactions. One the cubic lattice,
  results for antiferromagnetic interactions can be deduced from
  results using ferromagnetic couplings as the two cases can be mapped
  onto each other by a simple rotation of the pseudo-spins $\mathbf{T}$ on one sub-lattice
  \cite{Rynbach}.

For a long time, answering the question whether the EgM supports an
orbital-ordered low-temperature phase, indicated by a local order
parameter $\langle \mathbf{T} \rangle >0$, has been difficult due to
the presence of a sub-extensive ground state degeneracy. However, a
later rigorous analysis ~\cite{nussinov-2004-6,biskup_3DEG} showed
that the ground state degeneracy is lifted at finite temperature by an
order by disorder mechanism and that the EgM can indeed order into six
discrete ordering directions, given by \be\mathbb{T}^o_n = \left(
\cos[n\ 2\pi/6], \sin[n\ 2\pi/6] \right), \ee with $n=0,\dots,5$. This
analytical prediction was subsequently verified using classical MC
simulations~\cite{tanaka:267204,Rynbach}, and at higher temperatures a
continuous phase transition to a disordered phase has been
found. Strikingly, no propositions concerning the universality class
of this prototypical finite-temperature phase transition were made up
until now, neither analytically nor numerically. Here, we will address
this question and try to explore whether the special anisotropic
interactions lead to new critical phenomena, as for instance found in
dimer models \cite{FAlet-dimer,CharrierAlet2010}, or whether more
conventional magnetic universality classes \cite{vicarireview}
describe the orbital-ordering transition.
\section{Simulation technique, observables, and boundary conditions} 
The classical Hamiltonian \eref{eqn:eg_model} is considered here on a
simple cubic lattice of side length $L$ and volume $N=L^3$. We perform
state-of-the-art MC simulations along the lines of
Refs.~\cite{wenzelQCMPRB,wenzelCM2010}, a key feature being the use of
parallel-tempering methods as (so far) no cluster-like updates exist.

In order to detect long-range orbital ordering with $\langle \mathbf{T} \rangle > 0$,  possible order parameters 
are \be
\label{eqn:orderparam1}
m=(1/N) \sqrt{( \sum_i T^z_i )^2 + ( \sum_i T^x_i)^2},
\ee
which is the standard XY-order parameter, or alternatively
\be
\label{eqn:orderparam2}
m=(1/N) \left|\sum_i T^z_i \right| + (1/N)\left|\sum_i T^x_i \right|.
\ee Other order parameters are possible and were used
\cite{Rynbach}. In the following we use Eq.~\eref{eqn:orderparam2} in
Sec.~\ref{sec:EgM} and Eq.~\eref{eqn:orderparam1} in
Sec.~\ref{sec:EgMCL} to check the independence of our final results on
the definition of the order-parameter. Moreover, the complementary
quantity \be
\label{eqn:D}
D=(1/N)\sqrt{\left( E_x- E_y\right)^2 + \left( E_y-E_z \right)^2 +
  \left( E_{z\phantom{y}}\!\!-E_x \right)^2}, \ee indicates a
directional ordering of the bond energies and was previously studied
in the compass model \cite{mishra:207201,wenzelQCMPRB,wenzelCM2010}.
Here, $E_{x|y|z}$ is the total bond-energy along the
${x|y|z}$-direction (e.g., $E_x= -\ J\sum_{i} {\boldsymbol{\tau}}_i^x
{\boldsymbol{\tau}}_{i+\mathbf{e}_x}^x$).

For the finite-size scaling study reported here, we shall further make
use of the following quantities
\begin{eqnarray}
\label{eqn:sus}
\chi&=N(\langle m^2 \rangle - \langle m \rangle^2),\\
\label{eqn:mlogdbeta}
m'_{\ln}&=\mathrm{d} \ln m /\mathrm{d}\beta = N \left(\frac{\langle me \rangle} {\langle m \rangle} - \langle m \rangle \langle e \rangle \right),\\
\label{eqn:mdbeta}
m'&=\mathrm{d} m /\mathrm{d}\beta = N \left( \langle me \rangle - \langle m \rangle\langle e \rangle \right),\\
B_m&=1-\langle m^4\rangle/ 3 \langle m^2 \rangle^2,
\end{eqnarray}
denoting the susceptibility, the derivative of the logarithm of the
order parameter, the derivative of the order-parameter, and the Binder
parameter, respectively. Corresponding definitions apply to the order
parameter $D$.
\begin{figure}
\centering
\includegraphics[width=\textwidth]{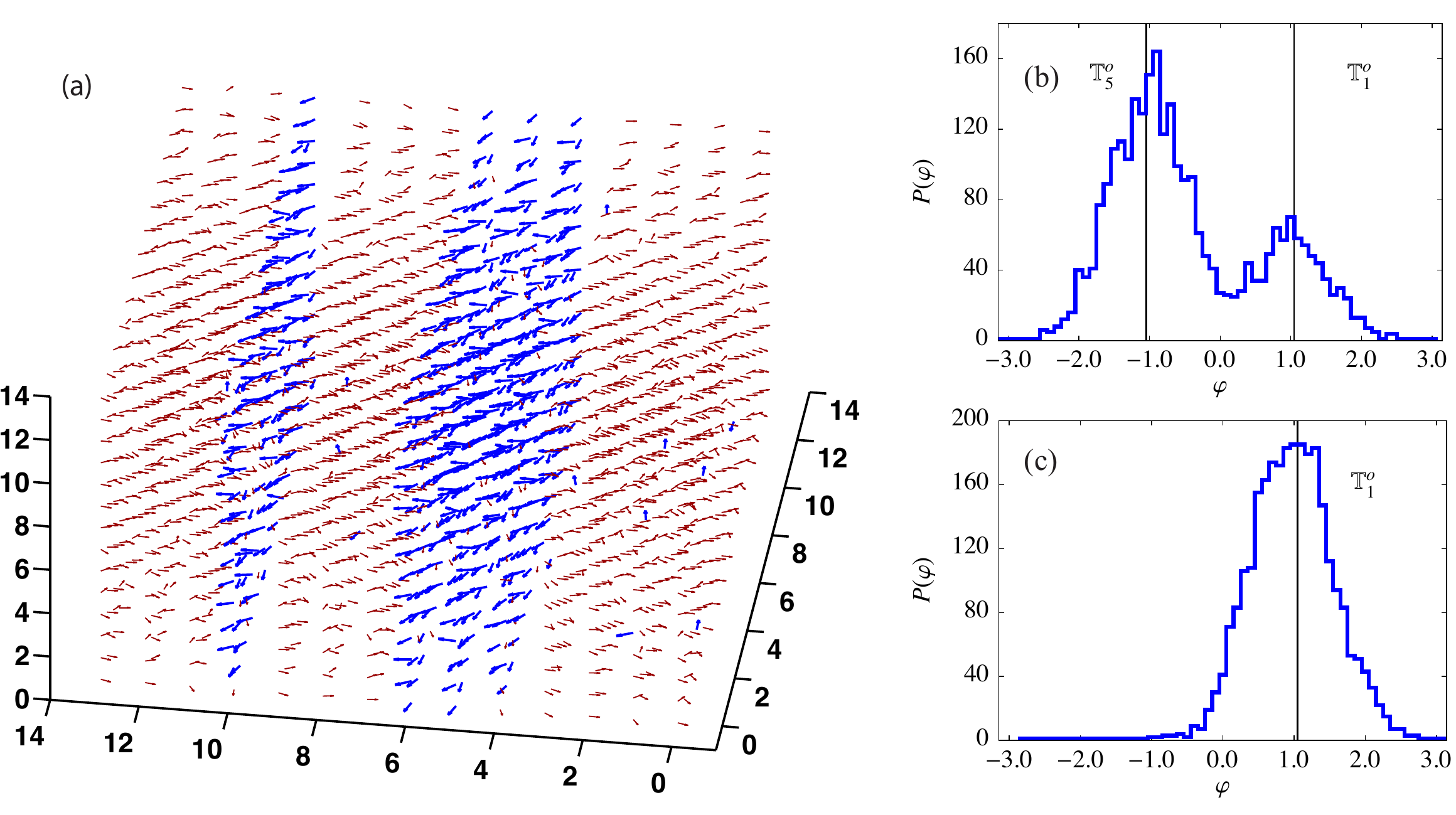}
\caption{\label{fig:hist1}Analysis of spin configurations in the
  ordered phase ($T=0.4$) after thermalization. (a) A typical
  configuration in real space for $L=14$ and periodic boundary
  conditions where the small arrows indicate the pseudo-spins
  $\mathbb{T}$. Spins are color-coded according to their major orientation.
 A coexistence of two phases is apparent and the phase
  boundaries are more or less planar, possibly due to the planar gauge
  symmetries at $T=0$. (b) A typical histogram of the distribution of
  spin angles $\varphi$ of one configuration snapshot for $L=14$ and periodic boundary conditions
  showing incomplete ordering with contribution from collective
  ordering angles $\mathbb{T}^o_1$ and $\mathbb{T}^o_5$. (c) Similar histogram using screw-periodic boundary
  conditions which largely favor just one collective spin orientation
  (here $\mathbb{T}^o_1$ as indicated by the vertical line).}
\end{figure}

A few preliminary MC test runs employing periodic boundary conditions
(PBC) clearly reproduce a signal of an ordered state at
low-temperature and of a thermal phase transition in accordance with
Ref.~\cite{Rynbach}. Moreover, by studying angular distribution
functions $P(\varphi)$ of the pseudo-spin angle $\varphi$ after
thermalization of a multitude of different initial configurations,
clear evidence for a six-fold degenerate ordering is found. However,
these initial runs also immediately reveal a couple of problems in the
simulations, the most evident being the presence of incomplete
ordering \emph{on small lattice sizes}. This incomplete ordering is
especially apparent by looking at a typical spin configuration
[Fig.~\ref{fig:hist1}(a)] or at angular distributions of the spins
(for one typical configuration) in a histogram
[Fig.~\ref{fig:hist1}(b)] which show coexistence of ordering regions
of different ordering angle $\mathbb{T}^o_n$. Such behavior is most
likely due to the presence of the (gauge-like) planar reflection
symmetries at $T=0$ \cite{nussinov-2004-6} which are also responsible
for the large ground-state degeneracy. For small system sizes, these
reflections are still not too unfavorable energetically even at
finite-temperatures. This problem could in principle be overcome by
using an alternative order parameter like the one of
Ref.~\cite{Rynbach} which is essentially insensitive to such ordering
metastabilities. However, we additionally find that there are rather
large finite-size corrections in any sort of scaling analysis on
periodic boundary conditions.

Previously, we have shown for the 2D compass model that under the
presence of gauge-like symmetries so-called screw-periodic boundary
conditions (SBC) can be very favorable \cite{wenzelCM2010}. Indeed,
SBC turn out to be very useful here as well: They naturally suppress
metastable regions by gluing together different planes thereby
favoring true collective ordering (visible in the histogram of
Fig.~\ref{fig:hist1}(c)). Second and more importantly, they in
principle allow to tune finite-size effects via the "screw parameter''
$S$. For points on the cubic lattice with coordinates $(x,y,z)$, SBC
can be defined by
\begin{equation}
\label{eq:typeASBC}
N_x(x,y,z)=\cases{(x+1,y,z) & if \quad $x<L-1$ \\ (0, [y+S]\,\mathrm{mod}\, L,z) & if\quad $x=L-1$,}
\end{equation}
where $N_x(x,y,z)$ denotes the nearest neighbor of $(x,y,z)$ in
x-direction.  This is one possible generalization of the definition
given in Ref.~\cite{wenzelCM2010}. A cyclic permutation is understood
for the other cases (going in $y$ and $z$-direction). Here, we report
results using $S=L/2$ which we empirically find to minimize
finite-size effects.

\section{\label{sec:EgM}Monte Carlo results and finite-size scaling for the $e_g$ model} 
\begin{figure}
\centering
\includegraphics[width=\textwidth]{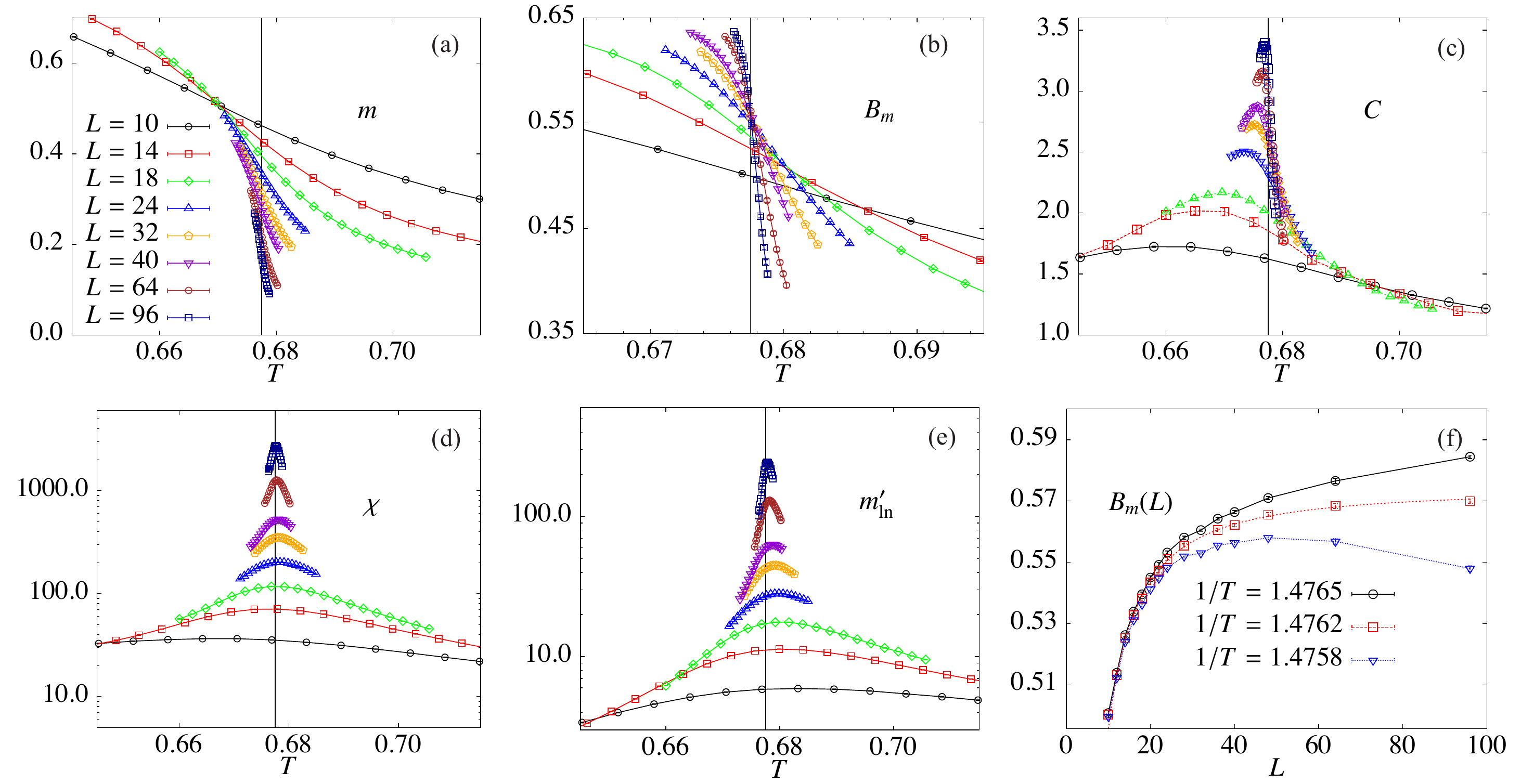}
\caption{Monte Carlo results for the {\em $e_g$ model}
  close to the phase transition: (a) The order parameter $m$, (b) the
  associated Binder cumulant $B_m$, (c) the heat-capacity $C$, (d) the
  susceptibility $\chi$, and (e) $m'_{\ln}$ as a function of
  temperature $T$ for different linear system sizes $L$. The vertical
  line indicates the location of the critical temperature $T_c$,
  obtained for example by an analysis of the finite-size scaling of
  $B_m(T_c)$ according to Eq.~\eref{eqn:BvsL} in (f).}
\label{fig:overview}
\end{figure}
We start by presenting numerical results for the
EgM~\eref{eqn:eg_model} with SBC on a couple of lattice sizes
$L=8,\dots,96$. To obtain the reported accuracy, we collected at least
$10^6$
or more independent MC measurements per data
point. Figure~\ref{fig:overview} displays some pertinent data for the
magnetization $m$ [using definition \eref{eqn:orderparam2}], the
Binder parameter $B_m$, the heat-capacity $C$, the susceptibility
$\chi$, and for $m'_{\ln}$ as a function of temperature $T$.  All
observables indicate a continuous phase transition at about
$T_c\approx 0.677$, in agreement with earlier PBC
estimates~\cite{tanaka:267204,Rynbach}. A first precise estimate of
$T_c$ can be obtained using the fact that $B_m(L)$ possesses only
corrections to scaling at the critical point, \be\label{eqn:BvsL} B_m(L)=B_m^\star + cL^{-\omega},\ee with
$\omega$ being the correction exponent. Figure~\ref{fig:overview}(f) shows
that this scaling is very well satisfied for $T_c=0.6775$, and an
\emph{effective} $\omega\approx 1.4$ with a large constant $c$.
\begin{figure}
\centering
\includegraphics[width=\textwidth]{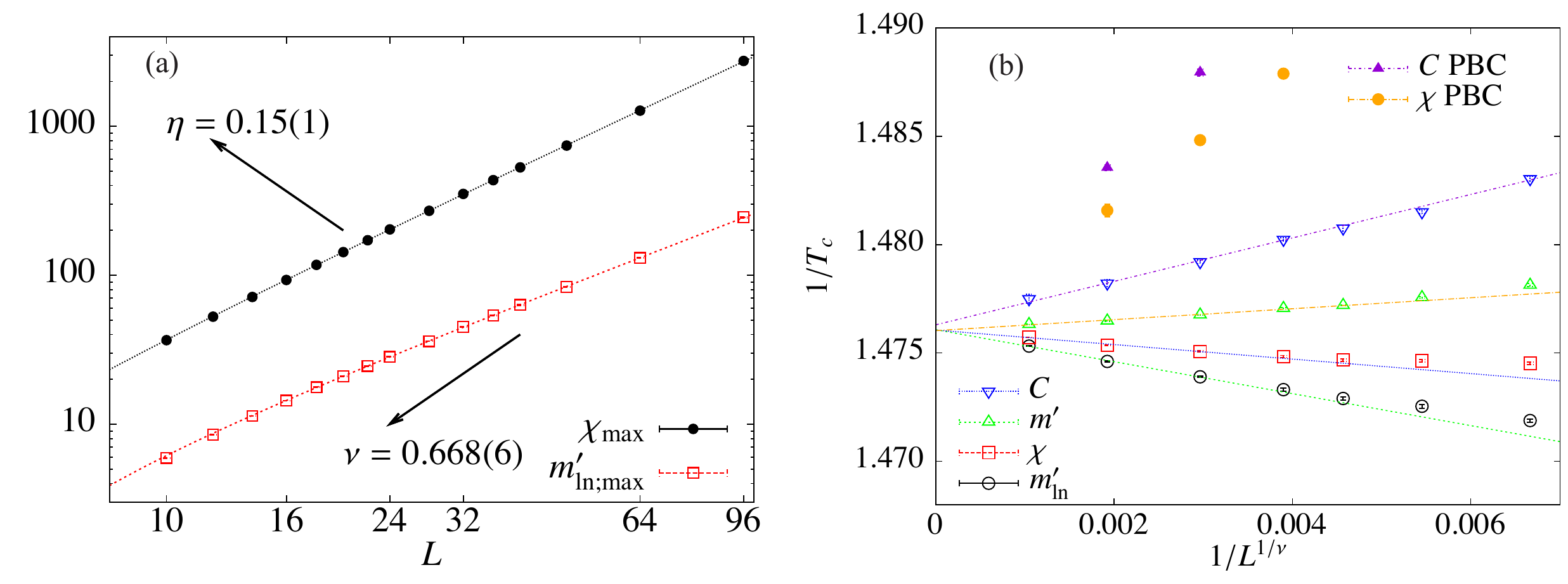}
\caption{\label{fig:FFS1}Finite-size scaling in the {\em $e_g$ model}: (a) Plot of $\chi_{\max}$ and $m'_{\ln;\max}$ versus $L$ in  a double logarithmic scale. Estimates for $\nu$ and $\eta$ where obtained from a finite-size study using Eq.~\eref{eq:ffs1} and Eq.~\eref{eq:ffs2}, taking into account corrections to scaling. The dashed lines are the corresponding fit curves. (b) }
\end{figure}
Based on this, we now perform a finite-size scaling study to obtain
the critical exponents. Here, we concentrate primarily on the
correlation length exponent $\nu$ describing the divergence of the
correlation length close to the critical point \be\xi\sim
\left|T-T_c\right|^{-\nu},\ee as well as the exponent $\eta$ governing
the decay of the spin-spin correlation function \be G(r)\sim
r^{-d+2-\eta} \ee at the critical point.  These exponents are
determined using $m'_{\ln;\max}=\max\{m'_{\ln}\}$ and the maximum of
the susceptibility, $\chi_{\max} = \max\{\chi\}$, which scale with
system size $L$ as
\begin{eqnarray}
\label{eq:ffs1}
m'_{\ln;\max}  \sim L^{1/\nu}(1+c_{m'} L^{-\omega}),\\
\label{eq:ffs2}
\chi_{\max} \sim L^{2-\eta}(1+c_{\chi} L^{-\omega}).
\end{eqnarray}
Using the effective correction exponent $\omega$ obtained above based
on the Binder cumulant, the data fits very well to Eq.~\eref{eq:ffs1}
yielding our estimate \be\nu=0.668(6)\ee for the correlation length
exponent, see Fig.~\ref{fig:FFS1}, which is roughly the same value as
that of the universality class of the 3D XY model with
$\nu_\mathrm{XY}=0.671$ \cite{hasenbuschXY,CampostriniXY2006}.
However, an analogous analysis of the order parameter correlations at
criticality yields \be\eta=0.15(1)\ee and provides strong
evidence for a universality class {\em distinct} from the 3D XY class,
which would yield a substantially smaller $\eta_\mathrm{XY}
\approx0.038$~\cite{vicarireview,hasenbuschXY}. 
Our main results for the critical exponents have been
reconfirmed by us using a slightly different but complementary analysis
(using ``running exponents''), without making use of $\omega$ \cite{WenzelEg2011}.

Having found $\nu$, one can return once more to the question of the
critical temperature which we want to obtain this time from the
scaling of pseudo-critical temperatures $T_c(L)$, defined from the
location of the peaks of the heat-capacity and of quantities defined
in Eqs.~\eref{eqn:sus},\eref{eqn:mlogdbeta}, and \eref{eqn:mdbeta}.
Those pseudo-critical temperatures $T_c(L)$ should scale according to
\be\label{eqn:scaling:Tc} T_c(L)=T_c + cL^{-1/\nu}(1+\cdots).\ee
Figure~\ref{fig:FFS1}(b) shows that such scaling is roughly satisfied
for the largest system sizes and that all quantities converge to a
unique $T_c$.  We give our final estimate as \be T_c=0.6775(1) \ee
which is the mean of all extrapolations.  This result is almost
insensitive to slight changes in the exponent $\nu$ within the error
bar.  Moreover, similar data obtained on periodic boundary conditions
[see Fig.~\ref{fig:FFS1}(b)] converge to the same critical point but
with evidently much larger finite-size effects, re-justifying the use
of screw-periodic boundary conditions.

We remark that other critical exponents, like the exponent $\alpha$ for the specific heat, have been
studied in a similar fashion. Our analysis yields $\alpha\approx 0$, which is in
agreement with the usual hyper-scaling relation.

\section{\label{sec:EgMCL}Monte Carlo results and finite-size scaling for the $e_g$-clock model} 
One might wonder whether the continuous nature of the orbital degrees
of freedom $\mathbf{T}$ is necessary for the critical properties
found. To address this question, let us consider here a naturally
discretized version of Hamiltonian~\eref{eqn:eg_model} -- one in which
the vectors $\mathbf{T}$ can only point along the six $\mathbb{T}^o$
ordering directions introduced above:
\begin{equation}
\label{egclock:model}
\mathcal{H}_{e_g}^\circledast= -\ J  \sum_{i,\alpha}  E^{\alpha}(n_i, n_{i+\mathbf{e}_\alpha})\,.
\end{equation}
Here, $E^{\alpha}(n_i,n_j)$ is the bond energy matrix along the bond
direction $\alpha$ and $n = 0,\ldots, 5$ denote the six discrete
onsite states $\mathbb{T}^o_{n}$. 
To be explicit, the following form of these matrices is easily obtained.
\begin{equation}
E^x(n,n^\prime)=\frac{1}{4} \pmatrix{ -4 & -2 & 2 & 4 & 2 & -2 \cr -2 & -1 & 1 & 2 & 1 & -1 \cr 2 & 1 & -1 & -2 & -1 & 1 \cr 4 & 2 & -2 & -4 & -2 & 2 \cr 2 & 1 & -1 & -2 & -1 & 1 \cr -2 & -1 & 1  & 2 & 1  & -1 }
\end{equation}
\begin{equation}
E^y(n,n^\prime)=\frac{1}{4} \pmatrix{ -1 & -2 & -1 & 1 & 2 & 1 \cr -2 & -4 & -2 & 2 & 4 & 2 \cr -1 & -2 & -1 & 1 & 2 & 1 \cr 1 & 2 & 1 & -1 & -2 & -1 \cr 2 & 4 & 2 & -2 & -4 & -2 \cr 1 & 2 & 1  & -1 & -2  & -2 }
\end{equation}
\begin{equation}
E^z(n,n^\prime)=\frac{1}{4} \pmatrix{ -1 & 1 & 2 & 1 & -1 & -2 
\cr 1 & -1 & -2 & -1 & 1 & 2
\cr 2 & -2 & -4 & -2 & 2 & 4 
\cr 1 & -1 & -2 & -1 & 1 & 2
\cr -1 & 1 & 2 & 1 & -1 & -2
\cr -2 & 2 & 4  & 2 & -2  & -4 }
\end{equation}
Note that via the above bond matrices, we can
introduce an interpolation between the EgCLM and the
three-state Potts compass model
\cite{Wenzel3DCM,mishra:207201} by multiplying all but the
matrix elements equal to $-1$ with a factor $\lambda\in
[0,1]$. Such interpolation could be useful to study the
crossover from a second-order phase transition to a first order transition found for the Potts compass
model \cite{Wenzel3DCM}.
The similarity of our model to the 6-state ($Z_6$) clock model \be
\mathcal{H}_{Z_6}=- J \sum_{\langle i,j \rangle}
\mathbb{T}^o_{n_i}\cdot\mathbb{T}^o_{n_j}\ee serves as a motivation to term
$\mathcal{H}_{e_g}^\circledast$ the \emph{$e_g$-clock model} (EgCLM)~\cite{WenzelEg2011}.

In addition to re-investigating critical exponents, we now also analyze 
the directional order parameter $D$ as introduced in
Eq.~\eref{eqn:D}. In an orbitally ordered state characterized by a
finite $m$, $D$ is also finite, however the converse is not true. An
illustrative example is given by the 2D compass model, where a
gauge-like freedom forbids orbital ordering
altogether~\cite{nussinovpip}, while $D$ orders at finite
temperature~\cite{mishra:207201,wenzelQCMPRB,wenzelCM2010}.
\begin{figure}
\centering
\includegraphics[width=\textwidth]{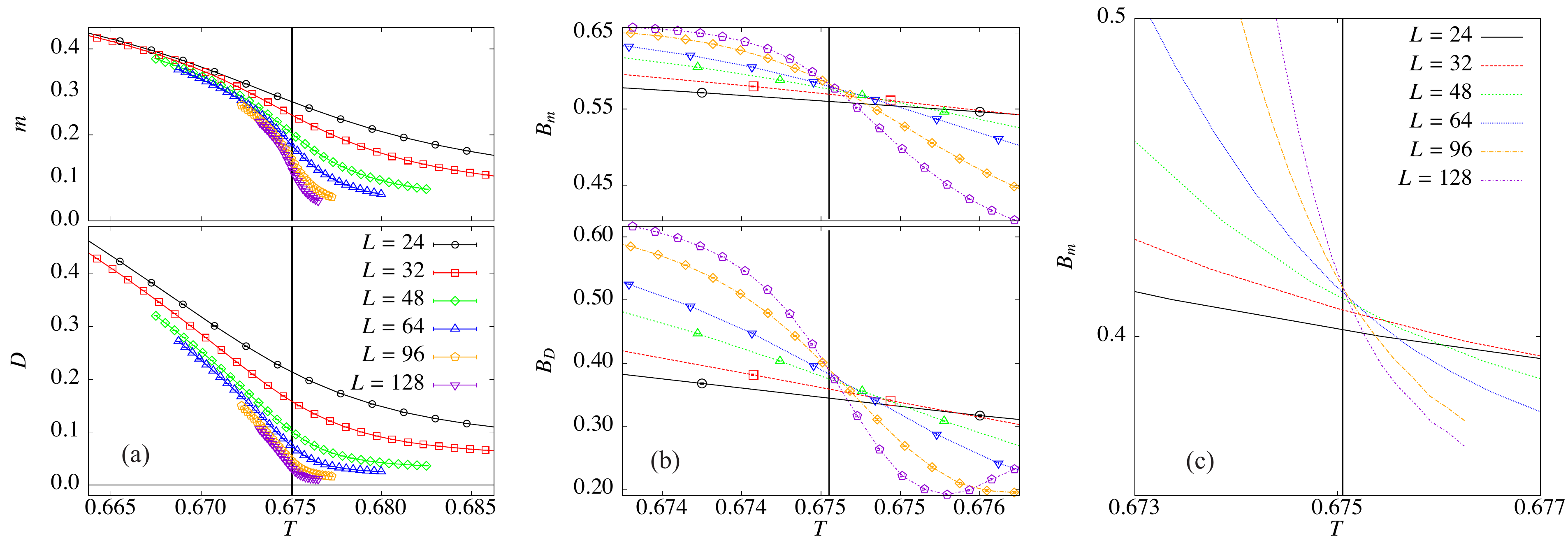}
\caption{\label{fig:EgCLM:D}Monte Carlo data for the
  {\em $e_g$-clock model}: (a) Orbital order parameter $m(T)$ (upper panel) and directional order parameter $D(T)$ (lower panel)
for different linear system sizes $L$. Note that both order parameters
become finite below a common $T_c$ (indicated by the vertical
line). (b) Binder parameters $B_m$ and $B_D$, and (c) the Binder
parameter $B_m$ as obtained from simulations on periodic boundary conditions.}
\end{figure}
Due to its discrete nature, we were able to study larger systems of up
to $L=128$ without much difficulty in our Monte Carlo runs. In
addition, this allowed us to perform a more systematic comparison
between periodic and screw-periodic boundary conditions.  In
Fig.~\ref{fig:EgCLM:D}, some of our resulting data is
presented. Clearly, both $m(T)$ and $D(T)$ show an ordering tendency
and both orbital ordering and directional ordering appear to set in at
about the same temperature [see Fig.~\ref{fig:EgCLM:D}(a)]. In order
to confirm the simultaneous onset we have in particular determined and
compared the respective Binder parameters $B_m$ and $B_D$, see
Fig.~\ref{fig:EgCLM:D}(b), indicating that both transitions take place
at a unique critical temperature $T_c=0.67505(3)$. This result rules
out a scenario of a directionally ordered, orbital-disordered
intermediate phase, and establishes a single transition from a high
temperature disordered phase to a low temperature orbitally ordered
phase. It is at this place appropriate to study and compare the Binder
parameter $B_m$ in more detail. At criticality, we note that
$B_m(T_c)\approx 0.589$ from the simulations on screw-periodic
boundary conditions. Interestingly, on periodic boundary conditions
the critical value of $B_m$ is much reduced to about $B_m(T_c)\approx
0.41$, as shown in Fig.~\ref{fig:EgCLM:D}(c). Hence, there is a huge
difference to the (weakly universal) value for $B_m\approx 0.586$
\cite{Toeroek} found for the standard 3D XY model \emph{on periodic
  boundary conditions}. Whether this large discrepancy is solely due
to a distinct universality class as found here, or partly based on the
directional nature of the ordered state remains to be investigated.

\begin{figure}[t]
\centering
\includegraphics[width=\textwidth]{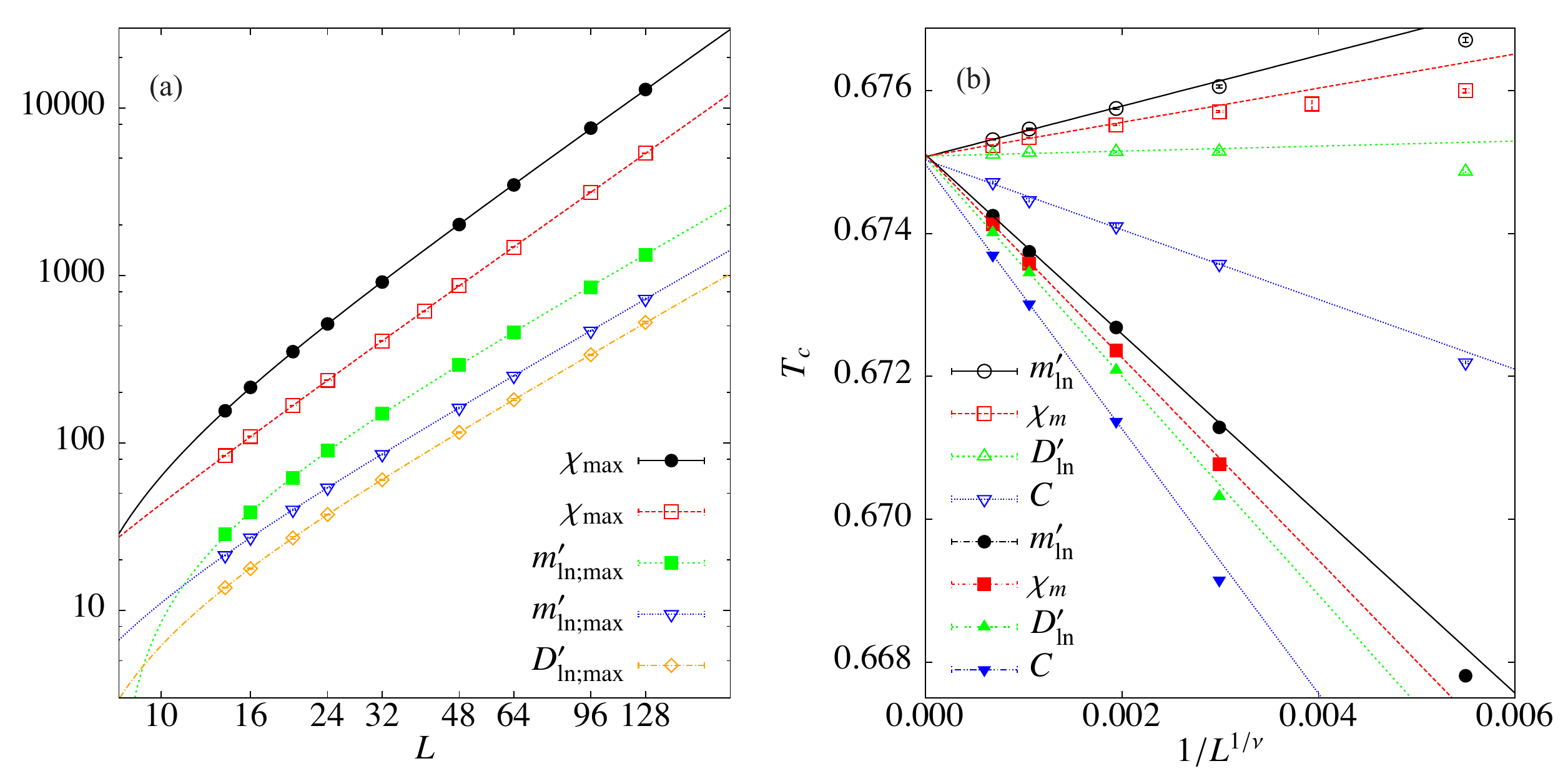}
\caption{\label{fig:scaling:EgClock}Finite-size
  scaling in the $e_g$-clock model. (a) Study of the exponents $\nu$
  and $\eta$ according to Eqs.~\eref{eq:ffs1} and
  \eref{eq:ffs2}. Open symbols are from simulation with
  screw-periodic boundary conditions while filled symbols are obtained
  using periodic boundary conditions. The latter suffer from larger
  corrections to scaling as evident from the curvature (in this
  log-log plot). (b) Scaling of the pseudo critical temperatures
  using $\nu=0.666$. Both the data from SBC (open symbols) and from
  PBC (filled symbols) nicely converge to the same critical point
  $T_c$.}
\end{figure}

We now present our results for the critical exponents in the EgCLM,
using the same scaling relations \eref{eq:ffs1} and \eref{eq:ffs2}
as before. Figure~\ref{fig:scaling:EgClock} shows the finite-size
scaling analysis and explicitly compares the scaling behavior for the two
different boundary conditions. Most importantly, we obtain the same
set of critical exponents as for the continuous $e_g$-model showing
that just the nature of the ordered state, 6-fold degeneracy plus
directional ordering, is relevant. Our most precise estimates
$\nu=0.666(5)$ and $\eta=0.15(1)$ are taken from SBC. Prior to
fitting, we have analyzed the critical Binder parameter yielding
roughly the same effective correction exponent as before.  The
corresponding data for PBC show much larger finite-size effects (strong
curvature in the log-log plot) but nevertheless give the same set of
exponents. To reinforce our findings and to check our procedure, we
also performed an analysis of the largely similar $Z_6$-clock model
yielding exactly the 3D XY values as predicted \cite{Jose_PRB,
  Hoveclock}.
Note that an analysis based on the order parameter $D$ instead
of $m$ leads to the same $\nu$ exponent (e.g. from $D'_{\ln;\max}$ in
Fig.~\ref{fig:scaling:EgClock}(a)), while the corresponding $\eta_D$
exponent is much larger $(\approx 1.4)$. This simply follows from the
assumption that $D$ has no intrinsic critical behavior, because then
$D$ is driven by $m$: $D \sim m^2$, resulting in an apparently
different $\eta$ value.

Last, we show a determination of the critical temperature from an
scaling analysis according to Eq.~\eref{eqn:scaling:Tc} as shown in
Fig.~\ref{fig:scaling:EgClock}(b). Again, multiple observables and boundary conditions
converge nicely to $T_c=0.67505(3)$.  Interestingly, the $T_c$ for the discrete
model is slightly smaller than that of the continuous
variant. This observation can tentatively be explained by the fact that ordered
phase is stabilized entropically.

\section{Emergence of a U(1) symmetry} 
\begin{figure}
\centering
\includegraphics[width=\textwidth]{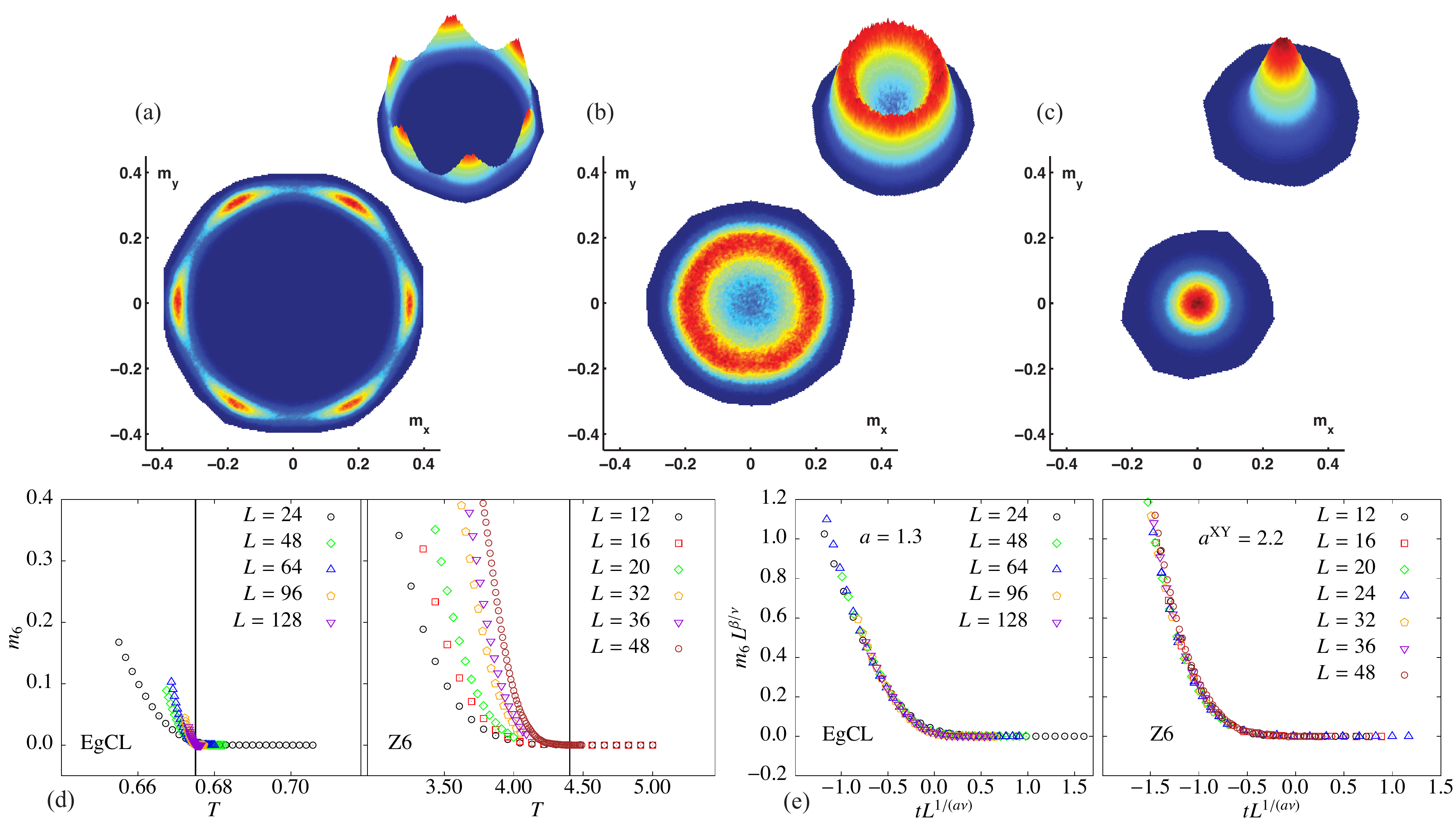}
\caption{\label{fig:U1scaling}(a-c) Distribution functions
  $P(m_x,m_y)$ (histograms) of the order parameter obtained for
  $e_g$-clock model for $L=64$ in the ordered phase $(T=0.67)$, just
  below $T_c$ ($T=0.675$), and in the disordered phase $(T=0.68)$. In
  each case a top view and a tilted three-dimensional view is given to
  illustrate the form of the distribution functions.  Values in the
  $z$-direction are shown on an arbitrary linear scale, which is
  therefore not shown. (d) The order parameter $m_6$ as defined in
  Eq.~\eref{eq:m6} for the $e_g$-clock model (EgCL) and the
  $Z_6$-clock model (Z6). For the EgCL, $m_6$ seems to become finite
  much more rapidly below $T_c$ (indicated by the vertical line) than
  in the Z6 case. (e) Collapse analysis of $m_6$: Best collapse
  parameters $a$ are indicated in the plot and differ clearly for the
  $e_g$-clock model and the $Z_6$-clock model.}
\end{figure}

In the last part of this presentation we study the order-parameter
distribution close to the critical point: In the ordered phase, a
six-fold degeneracy is present and it is interesting to see how this structure
is destroyed upon going into the disordered phase.  To this end, we
record histograms of the two-dimensional distribution $P(m_x,m_y)$,
where $m_x= (1/N)\sum_i T^z_i$ and $m_y =(1/N)\sum_i T^x_i$ (i.e., the
components of the vector order parameter $\langle \mathbf{T}
\rangle$). The same distribution in polar coordinates is denoted by
$P(r,\varphi)$. Fig.~\ref{fig:U1scaling}(a-c) shows a sequence of
histograms for a relatively large system $L=64$ obtained from
simulations of the EgCLM.  In the ordered phase, the six-fold
peak structure is recovered in the distribution function. However, at a
temperature just below $T_c$ a continuous and uniform distribution with a
finite radius shows up. This $U(1)$ symmetry of the orbital-order parameter is an
emergent symmetry not present in the Hamiltonian, a situation largely
reminiscent to the six-state $Z_6$ model, $Z_q$-perturbed XY
models or the antiferromagnetic three-state Potts model~\cite{Hoveclock,GottlobPotts,Heilmann96}. Emergent $U(1)$ symmetries of
discrete (i.e. dimer) order-parameters are also a major prediction of the theory of
deconfined critical points \cite{senthilDCScience,Sandvik2007}.

The emergent $U(1)$ symmetry at the critical point continues to govern the order parameter below 
a crossover length scale $\Lambda$ below  $T_c$ and this crossover scale is tied to the scaling
of the correlation length $\xi$ via~\cite{Jose_PRB,Blankschtein_PRB,Oshikawa_Zn,Louclock}
\be\Lambda \sim \xi^{a},\ee with an exponent $a$ 
depending solely on the universality class of the critical point, at least in the case of
$Z_q$-perturbed XY models~\cite{Oshikawa_Zn,Louclock}.


Hence, the value of $a$
defines another probe of unconventional critical behavior which we
want to address here. In order to obtain $a$, one considers a modified order parameter
\be
\label{eq:m6}
m_6= \int_0^1 d r \int_0^{2\pi}d\varphi r^2 P(r,\varphi)
\cos(6\varphi), \ee which is sensitive only to the sixfold symmetry
breaking, and which vanishes in the presence of a $U(1)$
symmetry~\cite{Louclock}.  Its finite-size scaling is thus influenced
by $\Lambda$ rather than $\xi$, i.e. $m_6$ will be finite whenever the
sixfold-structure is present in the histograms. In particular, it is
argued that close to criticality the following scaling relations \be
\label{eqn:ffsm6}
m_6 \sim L^{\beta/\nu} f(|t| L^{1/{a\nu}}), \quad m \sim L^{\beta/\nu}
g(|t|L^{1/\nu}) \ee hold, with $\beta$ being the critical exponent
associated with the order-parameter, $t$ the reduced temperature, and
$g$ and $f$ some scaling functions.  Relations \eref{eqn:ffsm6} allow
to extract $a$ via a typical collapse analysis of $m_6$, ideally using
a known value for $\beta/\nu$.  We have performed this analysis both
for the EgCLM and the $Z_6$ model as they have very similar ordered
states, see Fig~\ref{fig:U1scaling}(d,e). In the case of
the $Z_6$ model, we find $a^{XY}\approx 2.2$ (using $\beta/\nu=0.518$
for the 3D XY model). This result is consistent with the result of
Ref.~\cite{Louclock} but almost a factor two larger than the value
$a\approx 1.3$ that we obtain for the EgCLM (using
$\beta/\nu=(1+\eta)/2\approx 0.575$ for the 3D EG model). This result
constitutes therefore further support for the unconventional critical
behavior of the $120^\circ$ model found in other quantities.

\section{Conclusions}
We have studied the critical properties of the finite-temperature
ordering transitions in the $e_g$ or $120^\circ$ model which plays a
prototypical role in the study of collective effects resulting from
orbital-degeneracy.  Our systematic study points towards a distinct
universality class for orbital-ordering, different from the standard
(magnetic universality) classes we have encountered so far.  Next to
analyzing the original $e_g$ model, a discrete variant (the
$e_g$-clock model) was defined and found to exhibit the same critical
properties.  In comparison to magnetic universality classes,
unconventional critical properties are most apparent in the critical
exponent $\eta$, describing the critical correlation function, and in
the scaling of the length-scale $\Lambda$ related to the emergent
$U(1)$ symmetry of the order parameter at the critical point. Our work
provides a possible explanation of the unconventional observations
made in the presence of impurities in the $e_g$ model
\cite{tanaka:267204}.  Further theoretical work will be required to
shed light on our findings and to understand in more detail the
peculiar effects of the coupling of real space and order parameter
space~\cite{vicarireview,Nattermann1975}, which are at work in the
$120^\circ$ model. 
Recently, (artificially engineered) orbital systems
became available in solids
\cite{vandenBrink_NJP,Jackeli2009,Chaloupka2010} which gives promising
hope that the peculiar critical properties uncovered in the present
work can be further explored experimentally.



\ack
We thank K.~Binder, M.~Hasenbusch, G.~Misguich, R.~Moessner, M.~Oshikawa, and
S.~Trebst for useful discussions. SW thanks F.~Mila for discussions
and support. The simulations have been performed on the PKS-AIMS
cluster at the MPG RZ Garching and on the Callisto cluster at EPF
Lausanne.

\vspace{1cm}

\bibliographystyle{iopart-num}
\bibliography{literature}

\providecommand{\newblock}{}
\begin{thebibliography}{10}
\expandafter\ifx\csname url\endcsname\relax
  \def\url#1{{\tt #1}}\fi
\expandafter\ifx\csname urlprefix\endcsname\relax\def\urlprefix{URL }\fi
\providecommand{\eprint}[2][]{\url{#2}}

\bibitem{TokuraScienceReview}
Tokura Y and Nagaosa N 2000 {\em Science\/} {\bf 288} 462

\bibitem{KhomskiiJPhysA2003}
{Khomskii} D and {Mostovoy} M 2003 {\em J. Phys. A: Math. and Gen.\/} {\bf 36}
  9197

\bibitem{vandenBrink_NJP}
van~den Brink J 2004 {\em New J. Phys.\/} {\bf 6} 201

\bibitem{doucot:024505}
Dou\c{c}ot B, Feigel'man M, Ioffe L and Ioselevich A 2005 {\em Phys. Rev. B\/}
  {\bf 71} 024505

\bibitem{WenzelEg2011}
Wenzel S and L\"auchli A~M 2011 {\em Phys. Rev. Lett.\/} {\bf 106} 197201

\bibitem{Wenzel3DCM}
Wenzel S and L\"auchli A~M 2011 {\em unpublished\/}

\bibitem{Rynbach}
van Rynbach A, Todo S and Trebst S 2010 {\em Phys. Rev. Lett.\/} {\bf 105}
  146402

\bibitem{nussinov-2004-6}
Nussinov Z, Biskup M, Chayes L and {J~van den Brink} 2004 {\em Europhys.
  Lett.\/} {\bf 6} 990

\bibitem{biskup_3DEG}
Biskup M, Chayes L and Nussinov Z 2005 {\em Commun. Math. Phys.\/} {\bf 255}
  253

\bibitem{tanaka:267204}
Tanaka T, Matsumoto M and Ishihara S 2005 {\em Phys.~Rev.~Lett.\/} {\bf 95}
  267204

\bibitem{FAlet-dimer}
Alet F, Misguich G, Pasquier V, Moessner R and Jacobsen J 2006 {\em Phys. Rev.
  Lett.\/} {\bf 97} 030403

\bibitem{CharrierAlet2010}
Charrier D and Alet F 2010 {\em Phys. Rev. B\/} {\bf 82} 014429

\bibitem{vicarireview}
Pelissetto A and Vicari E 2002 {\em Phys. Rep.\/} {\bf 368} 549

\bibitem{wenzelQCMPRB}
Wenzel S and Janke W 2008 {\em Phys. Rev. B\/} {\bf 78} 064402

\bibitem{wenzelCM2010}
{Wenzel} S, {Janke} W and {L{\"a}uchli} A~M 2010 {\em Phys. Rev. E\/} {\bf 81}
  066702

\bibitem{mishra:207201}
Mishra A, Ma M, Zhang F~C, Guertler S, Tang L~H and Wan S 2004 {\em Phys.
  Rev.~Lett.\/} {\bf 93} 207201

\bibitem{hasenbuschXY}
Campostrini M, Hasenbusch M, Pelissetto A, Rossi P and Vicari E 2001 {\em Phys.
  Rev. B\/} {\bf 63} 214503

\bibitem{CampostriniXY2006}
Campostrini M, Hasenbusch M, Pelissetto A and Vicari E 2006 {\em Phys. Rev.
  B\/} {\bf 74} 144506

\bibitem{nussinovpip}
Nussinov Z and Fradkin E 2005 {\em Phys. Rev. B\/} {\bf 71} 195120

\bibitem{Toeroek}
Hasenbusch M and T\"or\"ok T 1999 {\em J. Phys. A: Math. Gen.\/} {\bf 32} 6361

\bibitem{Jose_PRB}
Jos\'e J~V, Kadanoff L~P, Kirkpatrick S and Nelson D~R 1977 {\em Phys. Rev.
  B\/} {\bf 16} 1217--1241

\bibitem{Hoveclock}
Hove J and Sudb\o{} A 2003 {\em Phys. Rev. E\/} {\bf 68} 046107

\bibitem{GottlobPotts}
Gottlob A~P and Hasenbusch M 1994 {\em Physica A: Statistical Mechanics and its
  Applications\/} {\bf 210} 217 -- 236 ISSN 0378-4371

\bibitem{Heilmann96}
Heilmann R, Wang J~S and Swendsen R 1996 {\em Phys. Rev. B\/} {\bf 53} 2210

\bibitem{senthilDCScience}
Senthil T, Vishwanath A, Balents L, Sachdev S and Fisher M 2004 {\em Science\/}
  {\bf 303} 1490

\bibitem{Sandvik2007}
Sandvik A~W 2007 {\em Phys. Rev. Lett.\/} {\bf 98} 227202

\bibitem{Blankschtein_PRB}
Blankschtein D, Ma M, Berker A~N, Grest G~S and Soukoulis C~M 1984 {\em Phys.
  Rev. B\/} {\bf 29} 5250--5252

\bibitem{Oshikawa_Zn}
Oshikawa M 2000 {\em Phys. Rev. B\/} {\bf 61} 3430--3434

\bibitem{Louclock}
Lou J, Sandvik A~W and Balents L 2007 {\em Phys. Rev. Lett.\/} {\bf 99} 207203

\bibitem{Nattermann1975}
Nattermann T and Trimper S 1975 {\em J. Phys. A: Math. Gen.\/} {\bf 8} 2000

\bibitem{Jackeli2009}
Jackeli G and Khaliullin G 2009 {\em Phys. Rev. Lett.\/} {\bf 102} 017205

\bibitem{Chaloupka2010}
Chaloupka J, Jackeli G and Khaliullin G 2010 {\em Phys. Rev. Lett.\/} {\bf 105}
  027204

\end{thebibliography}

\end{document}